\documentclass[%
 aip,
rsi,%
 amsmath,amssymb,
preprint,%
]{revtex4-1}

\usepackage{graphicx}
\usepackage{dcolumn}
\usepackage{bm}

\newcommand{\rmi}{\mathrm{i}}
\newcommand{\rme}{\mathrm{e}}
\newcommand{\rmd}{\mathrm{d}}
\newcommand{\sech}{\mathrm{sech}}
\newcommand{\csch}{\mathrm{csch}}

\begin{document}

\preprint{AIP/123-QED}

\title[Schr\"odinger formalism for a particle constrained to a surface in $\mathbb{R}_1^3$]{Schr\"odinger formalism for a particle constrained to a surface in $\mathbb{R}_1^3$}

\author{Renato Teixeira}
 \affiliation{Departamento de Matem\'atica, Universidade Federal Rural de Pernambuco, Recife, PE, 52171-900, Brazil}
 \affiliation{Departamento de Matem\'atica,  Universidade Federal 
de Pernambuco, Recife, PE, 50670-901, Brazil}
 
\author{Eduardo S. G. Leandro}%

\affiliation{Departamento de Matem\'atica,  Universidade Federal 
de Pernambuco, Recife, PE, 50670-901, Brazil}

\author{Luiz C. B. da Silva}
\altaffiliation[Part of this work was done while da Silva was a temporary lecturer at ]{Departamento de Matem\'atica,  Universidade Federal de Pernambuco, Recife, PE, 50670-901, Brazil}%
\affiliation{Department of Physics of Complex Systems, Weizmann Institute of Science, Rehovot 76100, Israel}

\author{Fernando Moraes}
 \email{fernando.jsmoraes@ufrpe.br}
\affiliation{Departamento de F\'{\i}sica, Universidade Federal Rural de Pernambuco, Recife, PE, 52171-900, Brazil}

\date{\today}

\begin{abstract}
In this work it is studied the Schr\"odinger equation for a non-relativistic particle restricted to move on a surface $S$ in a three-dimensional Minkowskian medium $\mathbb{R}_1^3$, i.e., the space $\mathbb{R}^3$ equipped with the metric $\text{diag}(-1,1,1)$. After establishing the consistency of the interpretative postulates for the new Schr\"odinger equation, namely the conservation of probability and the hermiticity of the new Hamiltonian built out of the Laplacian in $\mathbb{R}_1^3$, we investigate the confining potential formalism in the new effective geometry. Like in the well-known Euclidean case, it is found a geometry-induced potential acting on the dynamics $V_S = - \frac{\hbar^{2}}{2m} \left(\varepsilon H^2-K\right)$ which, besides the usual dependence on the mean ($H$) and Gaussian ($K$) curvatures of the surface, has the remarkable feature of a dependence on the signature of the induced metric of the surface: $\varepsilon= +1$ if the signature is $(-,+)$, and $\varepsilon=1$ if the signature is $(+,+)$. Applications to surfaces of revolution in $\mathbb{R}^3_1$ are examined, and we provide examples where the Schr\"odinger equation is exactly solvable. It is hoped that our formalism will prove useful in the modeling of novel materials such as hyperbolic metamaterials, which are characterized by a hyperbolic dispersion relation, in contrast to the usual spherical (elliptic) dispersion typically found in conventional materials.
\end{abstract}

\pacs{03.65.Ca, 42.70.-a, 02.40.Hw}
\keywords{Constrained dynamics, geometric potential, Minkowski space, surface of revolution, hyperbolic metamaterials}
\maketitle

\section{Introduction}

While  four-dimensional Minkowski space is  the gravity-free arena of special relativity, its three-dimensional version so far has been devoid of physical meaning, except as a lower dimensional toy model. The advent of hyperbolic metamaterials, though, has brought a physical realization of the 3D Minkowski space $\mathbb{R}_1^3$ in what concerns light propagation \cite{smolyaninov2013minkowski} and ballistic electrons motion \cite{figueiredo2016modeling}, for instance.  In short, hyperbolic metamaterials are highly anisotropic media which, in the case of electromagnetic propagation, combine metallic and insulating behaviors in different directions, leading to a hyperbolic dispersion relation \cite{poddubny2013hyperbolic}.  In the case of electronic metamaterials, the anisotropy may be interpreted as a result of the effective mass of ballistic electrons, which becomes a tensor and may have negative components \cite{dragoman2007metamaterials}. There is a very important analogy between the propagation of electromagnetic waves in dielectric media and  ballistic electrons in semiconductors \cite{henderson1991ballistic}  originating at the similarity between the  Helmholtz and the time-independent Schr\"odinger equations. Amazingly, this analogy survives even in the case of propagation along a surface, since both the quantum particle \cite{da1981quantum} and the electromagnetic wave \cite{willatzen2009electromagnetic} are subjected to the same effective geometry-induced potential, which comes from the extrinsic geometry of the surface.  

The purpose of this work is to extend to $\mathbb{R}_1^3$ the confining potential formalism developed by Jensen and Koppe \cite{JensenAndKoppeAP1971} and da Costa \cite{da1981quantum} for surfaces in ordinary Euclidean space, bearing in mind possible applications to media such as hyperbolic metamaterials. Even though our approach is via quantum mechanics, the optical analogy mentioned in the above paragraph allows for applications both in electromagnetic and electronic hyperbolic metamaterials. Incidentally, one of us (FM)  has recently  investigated optical propagation near topological defects in a hyperbolic metamaterial \cite{fumeron2015optics}. We emphasize that the 3D Minkowski geometry is suggested by the unusual dispersion relation characteristic of hyperbolic metamaterials, which makes possible a description in terms of an \textit{effective geometry} as an alternative to the conventional concept of an effective mass as recently done in Ref. \cite{Filgueiras2018AlternativeMetaconfinement}, where the background geometry is still Euclidean. 

In the early 1970's, Jensen and Koppe \cite{JensenAndKoppeAP1971} and later R. C. T.  da Costa, in  the 1980's \cite{da1981quantum}, published seminal works which  described the non-relativistic quantum motion of a particle confined to a surface by an external  potential acting along the direction normal to the surface. Considering a coordinate along this normal direction plus two curvilinear coordinates  on the surface, they verified that the wave equation is separable into a tangential and a normal part. The tangential equation includes a geometry-induced potential depending on the Gaussian and mean curvatures of the surface. Since the mean curvature is not preserved under isometries,  da Costa concluded that isometric surfaces could be associated with different Schr\"odinger equations. In a follow-up work \cite{da1982constraints}, da Costa generalized the confining potential formalism developed in \cite{da1981quantum} to a system consisting of an arbitrary number of particles, finding again the presence of geometry-induced potentials not invariant under isometries.  

Applications of the confining potential formalism abound in  two-dimensional systems, where geometry has been used to modify their electronic properties. For instance, in quantum waveguides \cite{exner2015,delCampoSR}, curved quantum wires \cite{exner1989bound}, ellipsoidal quantum dots \cite{cantele2000confined}, curved quantum layers \cite{duclos2001bound}, and corrugated semiconductor thin films \cite{ono2009tuning} to name a few. Recent works by our group explored the possibility of using da Costa-based geometric design to construct nanotubes with specific transport properties \cite{santos2016geometric} and investigated confining potential formalism in generalized cylinders  \cite{bastos2016quantum} and invariant surfaces \cite{daSilva2017quantumVgip}. The experimental verification of the  effects of the geometry-induced potential in a real physical system was realized by
measuring the high-resolution ultraviolet photoemission
spectra of a C$_{60}$ peanut-shaped polymer \cite{onoe2012observation}. In addition, the experimental realization of an optical analogue of the geometry-induced potential on a curve has also been reported \cite{szameit2010opticalAnalogue}.

In this work we follow the ideas of da Costa in order to investigate the behavior of a particle confined to a surface immersed in the 3D Minkowski space $\mathbb{R}^{3}_{1}$, i.e., in $\mathbb{R}^{3}$ endowed with the indefinite metric $\mbox{diag}(-1,1,1)$. We stress that we are studying $\mathbb{R}^3_1$ with the perspective of  metamaterial  applications, not as a toy model for Quantum Mechanics in spacetime of lower dimension. Thus the time-like coordinate, here denoted by $x_1$, is not ordinary time, which we consider an external parameter $t$. In the sequel, the terminology ``causal",  ``time-like", ``space-like'', or ``light-like" concerns the intrinsic geometry of $\mathbb{R}^3_1$ and bears no relation to physical time, i.e., following the current jargon, we employ this terminology only to classify objects in $\mathbb{R}_1^3$ according to their geometric properties. It  turns out that, for either a space-like or a time-like surface, the Schr\"odinger equation acquires a geometry-induced potential (as in the Euclidean case) which takes into account the causal character of the surface. We apply this result to surfaces of revolution, both for their intrinsic symmetry and for the wealth of possible surface types, a consequence of the anisotropy of $\mathbb{R}^{3}_{1}$. Depending on the causal characters of the profile curve and the plane that contains it, there is a choice of eight types of surfaces of revolution \cite{da2017moving}. We consider five of these, corresponding to the ones with a non-light-like axis of revolution: (i) space-like axis and time-like profile curve; (ii) space-like axis and space-like profile curve in a time-like plane; (iii) space-like axis and space-like profile curve in a space-like plane; (iv) time-like axis and time-like profile curve; and (v) time-like axis and space-like profile curve. As examples, we analyze the cases of one- and two-sheeted hyperboloids immersed in $\mathbb{R}^{3}_{1}$, whose rotation axis is time-like.  In both cases, the Schr\"odinger equation resembles a P\"oschl-Teller equation \cite{poschl1933bemerkungen}, yielding exact eigenvalues. 

This work is organized as follows. In section 2, we review some basic aspects of the geometry of surfaces  immersed in $\mathbb{R}^{3}_{1}$ and how to compute their Gaussian and mean curvatures. In section 3, we  establish the basic formalism for the Schr\"odinger equation in $\mathbb{R}_1^3$ and, in section 4, we follow our confining potential formalism in order to find the  Schr\"odinger equation for a particle constrained to move on a surface in $\mathbb{R}^{3}_{1}$. In section 5, we provide applications to surfaces of revolution and, in section 6, we present two examples where one can obtain exact solutions. Finally, in section 7, we present our conclusions. General references for Minkowski geometry are \cite{lopez,kuhnel2015differential,Oneil1983,thompson1996minkowski}.

\section{Surfaces in $\mathbb{R}^{3}_{1}$} 

In what follows we briefly review the basics of 3D Minkowski geometry, focusing on immersed surfaces and, in particular, surfaces of revolution. For more details we refer to \cite{lopez}. Minkowski space is naturally anisotropic and its 3D version, $\mathbb{R}^{3}_{1}$,  is just $\mathbb{R}^{3}$ endowed with the metric $L:=\left\langle\,\cdot\,,\,\cdot\,\right\rangle_{1}$ whose matrix is given by
$
[L_{ij}] = 
\mbox{diag}(-1,1,1).
$

The scalar product between two vectors  $x = (x_1,x_2,x_3)$ and $y=(y_1,y_2,y_3)$ in $\mathbb{R}_1^3$ is given by
$
\left\langle x,y\right\rangle_1=-x_1y_1+x_2y_2+x_3y_3, 
$
while the vector product is
$
x \times_{1} y = -(x_2 y_3 - x_3y_2)\hat{e}_1 - (x_1 y_3 - x_3y_1) \hat{e}_2 + (x_1 y_2 - x_2y_1)\hat{e}_3. 
$
The latter is defined from  the scalar triple product $\langle x\times_{1} y,z\rangle_1 = \det(x,y,z)$,
where $z=(z_1,z_2,z_3)$ and $(x,y,z)$ is the matrix whose columns are the coordinates of $x,\,y,$ and $z$.
Note that in this geometry the inner product of a vector by itself may be positive, negative, or null. This suggests the following classification for an arbitrary vector $v \in \mathbb{R}^3_1$:
    \emph{space-like}, if $\left\langle v,v\right\rangle_1 >0$ or $v=0$; \emph{time-like}, if $\left\langle v,v\right\rangle_1<0$; and \emph{light-like} if $\left\langle v,v\right\rangle_1 \,\,=0$ and $v\not=0$. Following the current usage found in the differential geometry literature, we shall refer to this classification as the ``causal character of a vector"  \cite{lopez,kuhnel2015differential,Oneil1983}, even though we are studying $\mathbb{R}^3_1$ with the perspective of metamaterials applications, not as a toy model of lower dimensional space-time. 

Following the same reasoning, curves and surfaces may have a causal character as well. For curves, the causal character is defined by the behavior of the tangent vector on all its points. A way of finding the causal character of a surface in $\mathbb{R}^3_1$ is by analyzing its induced metric, i.e., the restriction of the metric of  $\mathbb{R}^3_1$ on the tangent space $T_{p}S$ to the given surface $S$, at each point $p\in S$. If $S$ is parametrized by  $r(q_{1},q_{2})$ we have for the induced metric
\begin{eqnarray}
g_{ij}(p):= g\left(\frac{\partial r}{\partial q_{i}},\frac{\partial r}{\partial q_{j}}\right)\Big\vert_p = \left\langle \frac{\partial r}{\partial q_{i}},\frac{\partial r}{\partial q_{j}}\right\rangle_{1}\Big\vert_p. \label{indg}
\end{eqnarray}
This way, the surface is 
    \emph{space-like}, if $\forall\,p\in S$, $g(u,v)_{p}$ is  of signature $(+,+)$, 
    \emph{time-like}, if $\forall\,p\in S$, $g(u,v)_{p}$ is of signature  $(-,+)$, and 
    \emph{light-like}, if $\forall\,p\in S$, $g(u,v)_{p}$ is of signature  $(0,+)$. Another way of determining whether a surface is space-like, time-like or light-like, is by examination of its normal vector field, if it exists \cite{kuhnel2015differential}. Indeed, a surface is space-like (time-like) at $p$ if its normal $N$ at $p$ is time-like (space-like).  

Of course, there are curves and surfaces in $\mathbb{R}^3_1$ that do not fall in the above classification if the tangent (curves) or normal (surfaces) vectors have different causal characters at different points. In this work we focus specifically on time-like and space-like surfaces for the study of confined quantum particles. Among these, we choose surfaces of revolution for applications. 

For surfaces in  $\mathbb{R}_1^3$, the negative derivative of the unit vector field $N$ normal to the surface (Gauss map) is called Weingarten map, $A(v)=-\mathrm{d}N(v)$, whose matrix $[a_{ij}]$ is 
\begin{equation}\label{msff}
\left[\begin{array}{cc}
a_{11} & a_{12} \\
a_{21} & a_{22} \\
\end{array}\right] = -\varepsilon
\left[\begin{array}{cc}
h_{11} & h_{12} \\
h_{21} & h_{22} \\
\end{array} \right]
 \, \left[\begin{array}{cc}
g^{11} & g^{12} \\
g^{21} & g^{22} \\
\end{array} \right],
\end{equation}
where $h_{ij}=\langle N,\partial^2 r/\partial q_i\partial q_j\rangle_1$ are the coefficients of the \emph{second fundamental form}  and $g^{ij}$ the coefficients of the inverse of the metric, i.e., $g_{ik}\,g^{kj}=\delta_i^j$. The eigenvalues of this operator are the principal curvatures of the surface. Therefore, its trace and its determinant  define the mean and Gaussian curvatures of the surface, respectively. In $\mathbb{R}^3_1$, depending on the nature of the surface, the Weingarten operator might not be diagonalizable and, consequently, the Gaussian and mean curvatures may fail to be written as the product and the average of the principal curvatures. This is the case of some time-like surfaces, for instance.     Nevertheless, one can write the mean and Gaussian curvatures  as \cite{lopez}
\begin{equation}
H= \frac{\varepsilon}{2} \mbox{tr}(A) = \frac{\varepsilon}{2}\frac{g_{11}h_{22}-2g_{12}h_{12}+g_{22}h_{11}}{g_{11}g_{22}-(g_{12})^2} \label{H}
\end{equation}
and
\begin{equation}
K= \varepsilon \det(A)= \varepsilon\,\frac{h_{11}h_{22}-(h_{12})^2}{g_{11}g_{22}-(g_{12})^2}. \label{K}
\end{equation}
Here, $\langle N,N\rangle_1=\varepsilon=\pm1$ determines the causal character of $N$ and, consequently, of the surface, i.e., a space-like surface has a time-like normal vector  ($\varepsilon=- 1$) and a time-like surface has a space-like normal vector  ($\varepsilon= 1$).

\subsection{Surfaces of revolution in $\mathbb{R}^3_1$}

A \emph{rotation} in $\mathbb{R}^3_1$ is an isometry that leaves a certain straight line (the \emph{rotation axis}) pointwise fixed. Unlike the Euclidean case, where there is only one kind of rotation given by matrices like
\begin{equation}
\phi_{T}(q_{1}) = \left[\begin{array}{ccc}
1 & 0 & 0 \\
0 & \cos(q_{1}) & \sin(q_{1}) \\
0 & -\sin(q_{1}) & \cos(q_{1}) \\
\end{array} \right], \label{matrixT}
\end{equation}
which represents a rotation by an angle $q_1$ around the $x_1$-axis, there are more possibilities in Minkowski space due to its inherent anisotropy. There, if the rotation axis is time-like,  the above matrix applies. But if it is space-like,  we have a hyperbolic rotation (a boost in the context of relativity) around the $x_3$-axis like 
\begin{equation}
\phi_{S}(q_{1}) = \left[\begin{array}{ccc}
\cosh(q_{1}) & \sinh(q_{1}) & 0 \\
\sinh(q_{1}) & \cosh(q_{1}) & 0 \\
0 & 0 & 1
\end{array} \right]. \label{matrixS}
\end{equation}
In the former case, the points of the rotated curve  (generatrix) describe a circle, whereas in the latter they move along a hyperbola. Of course there are rotations around light-like axes \cite{da2017moving}, but in this work we consider only those surfaces of revolution with either space- or time-like axis.


At this point, we are ready to consider some specificities with respect to surfaces of revolution in Minkowski space.  In what follows we deal with five types of surfaces, classified in Table \ref{t_cronograma} according to the causal character of the respective axis of rotation, of the plane that contains the curve, and of the curve itself. 

\begin{table}[!htpb]
\centering
\begin{small} 
\begin{tabular}{|p{2.5cm}|p{2.5cm}|p{2.5cm}|p{2.5cm}|}
\hline
Axis & Plane  & Curve & Surface  \\ \hline \hline
time-like  & time-like  & time-like  & time-like  \\ \cline{3-4}
      &       & space-like & space-like \\ \hline 
space-like & time-like  & time-like  & time-like  \\ \cline{3-4}
      &       & space-like & space-like \\ \cline{2-4}
      & space-like & space-like & time-like \\ \hline
\end{tabular}
\vspace{10pt}
\end{small}
\caption{Causal character of surfaces of revolution in $\mathbb{R}_1^3$ as a function of the causal characters of their axes, planes, and profile curves.}
\label{t_cronograma}
\end{table}

\subsection{Surfaces with a space-like axis}
Here we consider three types of surfaces: the  ones generated by a profile curve (either space-like or time-like) in a time-like plane and the ones with space-like generatrix  in the space-like plane $x_2x_3$. We start with the generatrix in the plane $x_1x_3$. Let $\alpha(q_2)=(u(q_2),0,v(q_2))$ be its parametrization with $u>0$. It follows that the parametrization of the surface of revolution  is obtained from the application of \eqref{matrixT} to $\alpha(q_2)$:
\begin{equation*}
r(q_{1},q_{2}) := 
(u(q_{2})\cosh(q_{1}),u(q_{2})\sinh(q_{1}),v(q_{2})). 
\end{equation*}
The induced metric is obtained by feeding $r(q_{1},q_{2})$ to  \eqref{indg}:
$ 
\left[g_{ij}\right] =\mbox{diag}[u^{2},(v')^{2}-(u')^{2}]. 
$
If $\alpha$ is parametrized by its arc length, we have that  $\eta=\langle\alpha',\alpha'\rangle_1$= $(v')^{2} - (u')^{2}$, where $\eta$ = $1$ if the curve is space-like, and $\eta$=$-1$, if it is time-like. The metric matrix then takes the form 
\begin{equation}
\left[g_{ij}\right] = \left[\begin{array}{cc}
u^{2} & 0 \\
0 & \eta \\
\end{array} \right] .
\end{equation} 
The non-null coefficients of the Weingarten operator are
$
a_{11} = -v'/u$ and $ 
a_{22} = \eta(v''u'-v'u'').  
$

If the curve is in the space-like plane $x_2x_3$, then its parametrization is given by  $\alpha(q_{2}) = (0,u(q_{2}),v(q_{2}))$, $u>0$. Following the above steps, we find for this case
$ 
\left[g_{ij}\right] = \mbox{diag}[
-u^{2},\,(u')^{2}+(v')^{2}] .
$
Again, if $\alpha$ is parametrized by its arc length, then it can only be space-like. Thus
\begin{equation} 
\left[g_{ij}\right] = \left[\begin{array}{cc}
-u^{2} & 0 \\
0 & 1\\
\end{array} \right]
\end{equation}
and, therefore, $
a_{11} = -v'/u,\, 
a_{22} = (v'u''-v''u').  
$

\subsection{Surfaces with a time-like axis}
Now we consider the cases of either space-like or time-like  generatrices in  a time-like plane. By applying \eqref{matrixT} to the curve
 $\alpha(q_2)$ = $(u(q_2),0,v(q_2))$, $v>0$, in the plane $x_1x_3$, we get for the corresponding surface
\begin{equation}
r(q_{1},q_{2}) := 
{(u(q_{2}),v(q_{2})\sin(q_{1}),v(q_{2})\cos(q_{1})). }
\end{equation}
As a consequence, $ 
\left[g_{ij}\right] = \mbox{diag}[v^{2},(v')^{2}-(u')^{2}] 
$
and if $\alpha$ is either space-like or time-like and can be parametrized by arc length, we have that $ \langle \alpha',\alpha' \rangle_1 $ = $\eta$. Therefore the metric becomes 
\begin{equation}
\left[g_{ij}\right] = \left[\begin{array}{cc}
v^{2} & 0 \\
0 & \eta \\
\end{array} \right]
\end{equation}
and again by \eqref{msff}, the non-null Weingarten coefficients are
\begin{equation}\label{alTijA}
a_{11} = \dfrac{-u'}{v}, \, 
a_{22} = \eta(v''u'-v'u'').  
\end{equation}

Note that all surfaces of revolution here considered, the Weingarten operator is diagonal implying that $a_{11}$ and $a_{22}$ correspond to the principal curvatures of the surfaces. The induced metrics are diagonal as well ($g_{12}=0$). 

\section{Schr\"odinger equation in $\mathbb{R}^{3}_{1}$}

The Schr\"odinger equation in $\mathbb{R}^{3}_{1}$ can be obtained by just replacing the usual Laplace operator (in an otherwise Riemannian background) by the respective Laplacian of the Minkowski metric. The new Laplacian operator is often named \textit{d'Alembertian} and denoted by $\square^2$ (or just $\square$). Notice, however, that some important questions call for an appropriate answer, such as (i) Is the new Hamiltonian a Hermitian operator? and, consequently, (ii) Are its eigenvalues real? (iii) Can the solution of this new equation be interpreted probabilistically as in the usual Quantum Theory? We shall see in the following that the above questions have a positive answer and, consequently, the formal mathematical structure associated with the new Schr\"odinger equation in the effective geometry of  $\mathbb{R}^3_1$ can be borrowed from the usual one in $\mathbb{R}^3$.

By denoting the gradient of a function $\psi$ in the metric $\langle\cdot,\cdot\rangle_1$ by 
$
\nabla_1 \psi = (-\partial_x\psi,\partial_y\psi,\partial_z\psi)$,
the d'Alembertian operator, i.e., the Laplacian in $\mathbb{R}_1^3$, may be written as
\begin{equation}
\label{eq::dAlembOperator}
\square^2 \psi = \langle\nabla_1,\nabla_1\psi\rangle_1=-\partial_x^2\psi+\partial_y^2\psi+\partial_z^2\psi\,.
\end{equation}
Observe that we can also write $\mbox{div}(\nabla_1\psi)=\square^2\psi$, where $\mbox{div}(\cdot)$ is the divergence computed with respect to the Euclidean metric ($\nabla_1$ is still computed with $\langle\cdot,\cdot\rangle_1$). Thus, if we define the effective momentum operator in $\mathbb{R}_1^3$ to be $\hat{p}=-\rmi\,\hbar\nabla_1$, the respective Schr\"odinger equation reads
\begin{equation}
\label{eq::SchrodingerEqInMinkowski}
\rmi\,\hbar\frac{\partial\psi}{\partial t} = \left(\frac{\hat{p}\,^2}{2m}+V\right)\psi=-\frac{\hbar^2}{2m}\square^2\psi+V\psi\,,
\end{equation}
where $V$ is a real function representing a given potential.

Now let us introduce, in the space of square integrable functions $L^2(I\times\Omega)$, the inner product
\begin{equation}
(\phi,\psi) = \int_{I\times\Omega}\rmd^3\mathbf{x}\, \phi^*(t,\mathbf{x})\psi(t,\mathbf{x})\,,
\end{equation}
where $^*$ denotes complex conjugation, $I\subseteq\mathbb{R}$ is an interval, and $\Omega\subseteq\mathbb{R}^3$ is a domain whose boundary $\partial\Omega$ is orientable and its light-like points form a measure zero set (outside this set a unit normal $N$ can be properly defined). In addition, let us assume Dirichlet or Neumann boundary conditions in $\partial\Omega$, i.e., $\psi\equiv0$ or $\partial\psi/\partial N=\langle \nabla_1\psi,N\rangle_1\equiv0$ in $\partial \Omega$, respectively (if $\Omega$ extends to infinity we shall assume that the wave functions and its derivatives decay sufficiently fast to zero). Then, we have for $-\square^2$ the relation
\begin{eqnarray}
(\phi,-\square^2\psi) & = & -\int_{\Omega}\rmd^3\mathbf{x}\, \phi\,\mbox{div}(\nabla_1\psi)\nonumber\\
& = & -\int_{\Omega}\rmd^3\mathbf{x}\, \mbox{div}(\phi\,\nabla_1\psi)+\int_{\Omega}\rmd^3\mathbf{x}\, \langle\nabla_1\phi,\nabla_1\psi\rangle_1\nonumber\\
& = & \int_{\Omega}\rmd^3\mathbf{x}\, \langle\nabla_1\phi,\nabla_1\psi\rangle_1\nonumber\,,\label{eq::PhiInnerdAlembPsiAsAnIndefiniteForm}
\end{eqnarray}
where we used the divergence theorem in combination with the boundary conditions. From the equation above we deduce that (i) $-\square^2$ is a Hermitian operator in $L^2(I\times\Omega)$, (ii) the eigenvalues of $-\square^2$, if they exist, are real and, unlike the usual Laplacian, (iii) the eigenvalues may be negative, positive, or null since $b(\phi,\psi)=\int_{\Omega}\rmd^3\mathbf{x}\, \langle\nabla_1\phi,\nabla_1\psi\rangle_1$ is an indefinite bilinear form. Similar computations and conclusions are also valid for $-\frac{\hbar^2}{2m}\,\square^2+V$.  
\newline
\newline
\textit{Example:} Consider a particle in a box $[0,a]\times[0,b]\times[0,c]$. As can be easily verified, the solutions of $-\frac{\hbar^2}{2m}\square^2\psi=E\psi$ are
\begin{equation}
\left\{
\begin{array}{c}
\psi_{n_1,n_2,n_3}=\sin(\frac{n_1\pi}{a}x_1)\sin(\frac{n_2\pi}{b}x_2)\sin(\frac{n_3\pi}{c}x_3)\\[4pt]
E_{n_1,n_2,n_3}=\frac{\hbar^2}{2m}(-\frac{n_1^2\pi^2}{a^2}+\frac{n_2^2\pi^2}{b^2}+\frac{n_3^2\pi^2}{c^2})\\
\end{array}
\right..
\end{equation}
As expected, the dispersion relation is hyperbolic due to the negative effective mass in the $x_1$-direction. In contrast with the usual quantum mechanics, notice that the energy spectrum is not bounded from below ($E\to\pm\infty$) and, in addition, if the sides of the box are commensurate, there may appear energies which are infinitely degenerate: e.g., if $a=b=c$, then for all $n=1,2,\dots$ we have $E_{n,n,n}=0$.

Unbounded energies and infinitely degenerate states are not found for Laplacians in Riemannian geometry. Indeed, under appropriate conditions, in a Riemannian manifold $M$ we usually have: (i) the Laplacian $\Delta_M$ extends to a self-adjoint operator on $L^2(M)$; (ii) there exist infinitely many $L^2$-eigenvalues of $\Delta_M$; (iii) an eigenfunction of $\Delta_M$ is infinitely differentiable; (iv) each eigenspace of $\Delta_M$ is finite-dimensional; and (v) the set of $L^2$-eigenvalues is discrete in $\mathbb{R}$. The  third to fifth properties, however, may fail in the semi-Riemannian case \cite{Kobayashi2016Laplacian} (from the examples mentioned in \cite{Kobayashi2016Laplacian}, we see that the spectrum of the Laplacian in semi-Riemannian geometry is a meaningful concept.)
    
Finally, it is worth mentioning the existence of an alternative notion of indefinite Laplacian related to metamaterials in which the  electric permittivity and/or magnetic permeability are/is negative. In such cases, the domain $\Omega\subseteq\mathbb{R}^n$, $n\geq1$, is written as $\Omega=\Omega_+\cup\Omega_-$, with a smooth interface between $\Omega_{\pm}$, and the Laplacian $\mathcal{A}$ is \cite{BehrndtJAM2018}
\begin{equation}
\mathcal{A}(f)=\left\{
\begin{array}{c}
-(\nabla\cdot\nabla f)\vert_q,\mbox{ if }q\in\Omega_+\\
+(\nabla\cdot\nabla f)\vert_q,\mbox{ if }q\in\Omega_-\\
\end{array}
\right..\label{eq::AlternativeIndLaplacian}
\end{equation}
Notice that here the effective mass $m^*$ depends on the position: $m^*>0$ in $\Omega_+$ and $m^*<0$ in $\Omega_-$. On the other hand, hyperbolic metamaterials are characterized by a hyperbolic dispersion relation and, consequently, the effective mass should depend on the direction \cite{ShekharNanoC2014}. Anisotropic effective masses is a crucial feature for a modeling through an effective geometry, since one can conveniently choose an effective linear momentum, $\hat{\mathbf{p}}_{\text{\it eff}}=-\rmi\hbar\,\mbox{grad}\,$, as a result of an effective metric.

\subsection{Probability and current densities}

Let $\psi$ be a wave function, i.e.,  a solution of Eq. (\ref{eq::SchrodingerEqInMinkowski}). The probability density may be defined as
$\rho(t,\mathbf{x}) = \psi^*(t,\mathbf{x})\psi(t,\mathbf{x}).
$ 
Now, using (\ref{eq::SchrodingerEqInMinkowski}) and (\ref{eq::PhiInnerdAlembPsiAsAnIndefiniteForm}), we have
\begin{equation}
\int_{\Omega}\rmd^3\mathbf{x}\,\psi^*\frac{\partial\psi}{\partial t} =
-\frac{\rmi\hbar}{2m}\int_{\Omega}\rmd^3\mathbf{x}\,\left(\langle\nabla_1\psi,\nabla_1\psi\rangle_1+\frac{2mV}{\hbar^2}\vert\psi\vert^2\right).
\end{equation}
It follows that the  real part of the expression on the left-hand side vanishes, i.e., $\Re(\int_{\Omega}\,\psi^*\partial_t\psi)=0$. We now use the equation above to show that the probability $\int_{\Omega}\rho$ is conserved. Indeed,
\begin{equation}
\frac{\rmd}{\rmd t}\int_{\Omega}\rmd^3\mathbf{x}\,\psi^*\psi =2\,\Re\left(\int_{\Omega}\rmd^3\mathbf{x}\,\psi^*\frac{\partial\psi}{\partial t}\right)=0\,.
\end{equation}
In particular, the conservation of probability implies that there exists at most one solution of (\ref{eq::SchrodingerEqInMinkowski}) for a given initial condition $\psi(0,\mathbf{x})=\psi_0(\mathbf{x})$.

In addition, if $\psi$ is a wave function, the derivative of $\rho$ can be written as
\begin{equation}
\label{eq::DerivativeOfRho}
\frac{\partial \rho}{\partial t} = -\frac{\hbar}{2m\rmi}\langle\nabla_1,\psi^*\nabla_1\psi-\psi\nabla_1\psi^*\rangle_1\,.
\end{equation}
Introducing the current density $\mathbf{j}= \frac{\hbar}{2m\rmi}(\psi^*\nabla_1\psi-\psi\nabla_1\psi^*)\,,$ it follows that (\ref{eq::DerivativeOfRho}) is just the continuity equation describing the local conservation of the probability density $\rho$:
\begin{equation}
\label{eq::ContinuityEq}
\frac{\partial\rho}{\partial t}+\langle\nabla_1,\mathbf{j}\,\rangle_1=0\,\mbox{ or }\,\frac{\partial\rho}{\partial t}+\mbox{div}(\mathbf{j})=0\,.
\end{equation}

\section{Schr\"odinger equation for a particle confined to a surface in $\mathbb{R}^{3}_{1}$}

\begin{figure}[ht]
    \centering
    \includegraphics[width=0.4\linewidth]{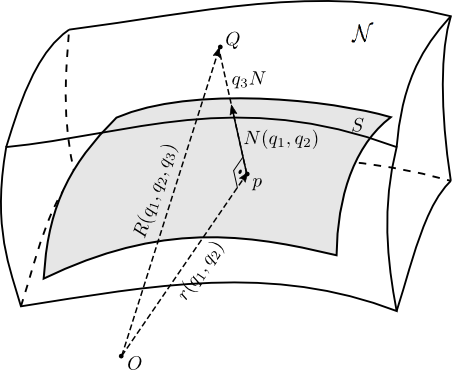}
    \caption{Coordinates in a tubular neighborhood $\mathcal{N}$ of $p\in S$. }
    \label{CdC}
\end{figure}

  Let $p=r(q_{1o},q_{2o})$ be a point in $S$ and $\mathcal{N}$ be a neighborhood of $p$ in $\mathbb{R}^{3}_{1}$. Following da Costa we endow the neighborhood $\mathcal{N}$ with an orthogonal coordinate system $q_1,q_2,q_3$, such that $q_1$ and $q_2$ are internal coordinates that parametrize the surface and $q_3$ is a coordinate along the surface's normal direction.  Then, the position of a point $(q_1,q_2,q_3)\in \mathcal{N}$ is given by
\begin{equation}\label{eqR}
R(q_{1},q_{2},q_{3}) = r(q_{1},q_{2}) + q_{3}N(q_{1},q_{2}),
\end{equation}
where 
$N(q_1,q_2)=(|\det{g}|)^{-\frac{1}{2}}\left(\partial_{q_1} r \times_1 \partial_{q_2} r\right)$ is a unit normal to the surface {at $(q_{1},q_{2})$}. See Fig. \ref{CdC} for a graphical representation of the coordinate system defined above.

Since $\langle N,N\rangle_1=\varepsilon=\pm1$, then $\langle\partial N/\partial q_{i},N \rangle_{1} = 0$. It follows that $\partial N/\partial q_{i}$  is in the tangent plane. So, 
\begin{equation}\label{eqdN}
\frac{\partial N}{\partial q_{i}}  =  \sum_{j=1}^2a_{ij} \frac{\partial r}{\partial q_{j}},
\end{equation}
where $a_{ij}$ are the coefficients of the Weingarten operator.

The induced metric in $\mathcal{N}$ is
$G_{ij} := \langle \partial R/\partial q_{i}, \partial R/\partial q_{j} \rangle_{1}$, $\,i,j=1,\,2,\,3$. For $i,j=1,2$ we obtain
\begin{equation}\label{Metri1}
G_{ij}    
= \displaystyle \sum_{k,l=1}^{2}\left(\delta_{ik} + q_{3} a_{ik}\right)\left(\delta_{jl} + q_{3} a_{jl}\right)g_{kl}, 
\end{equation}
and 
\begin{equation}\label{Metri2}
G_{i3}  = \left \langle \displaystyle \sum_{k=1}^{2}\left(\delta_{ik} + q_{3} a_{ik}\right) \frac{\partial r}{\partial q_{k}},N \right \rangle_{1} = 0,\,G_{33} =  \varepsilon.
\end{equation}

We are interested in the Schr\"odinger equation for a particle moving in $\mathcal{N}$. That is,
\begin{equation}\label{eqsch}
\rmi\hbar\frac{\partial \psi(x,t)}{\partial t}=-\frac{\hbar^{2}}{2m} \Delta_{G} \psi + V_{\lambda}(q_{3})\psi,
\end{equation}
where $\Delta_{G}$ is the Laplace operator associated with the metric $G$ and $V_{\lambda}$ is a generic confining potential acting along the normal coordinate $q_3$ such that
\begin{equation}\label{eq::DefViaLimVlambda}
\lim_{\lambda \rightarrow \infty} \varepsilon V_{\lambda}(q_{3})= 
\begin{cases}
\,0 & q_3=0,
       \\
    \infty & q_3 \neq 0.
       \\
\end{cases}
\end{equation}
The parameter $\lambda$ keeps track of the strength of the confinement. We emphasize that  $V_{\lambda}$ needed to be adjusted to conform with the causal character of the surface by multiplication by $\varepsilon$. In addition, to formally achieve the limit expressed in equation (\ref{eq::DefViaLimVlambda}) above, we may consider a sequence $\{ V_{\lambda} \}_{\lambda\geq0}$ of potentials corresponding to homogeneous boundary conditions along two neighboring surfaces equidistant from $S$ \cite{JensenAndKoppeAP1971}, say at a distance $\delta=1/\lambda$. In other words, for each $\lambda$, $V_{\lambda}(q)$ vanishes if $\mbox{dist}(q,S)\leq \delta$ and explodes, i.e., $\varepsilon V_{\lambda}=+\infty$, if otherwise. This is a crucial issue since the effective confined dynamics is sensitive to the way the confinement is formally achieved \cite{WangEtAlPRA2018}, as well as the possibility of decoupling the low energy tangential degrees of freedom from the high energy normal ones \cite{WachsmuthPRA2010,SchmelcherPRA2014}.

With the usual expression for the Laplacian in an $n$-dimensional  semi-Riemannian manifold endowed with a generic metric $g$,
\begin{equation}
\Delta_{g}f = \displaystyle\sum_{i,j=1}^{n}\frac{1}{\sqrt{|\det(g)|}}\frac{\partial}{\partial q_{i}}\left(\sqrt{|\det(g)|}\,g^{ij}\frac{\partial f}{\partial q_{j}}\right),
\end{equation}
where $g^{ij}$ are the coefficients of the inverse $[g_{ij}]^{-1}$, equation (\ref{eqsch}) becomes
\begin{equation}
-\frac{\hbar^{2}}{2m}\left\{ D(q)\psi 
+ \varepsilon  \left[\frac{\partial}{\partial q_{3}}\left(\ln |G|^{\frac{1}{2}} \right)\frac{\partial \psi}{\partial q_{3}} + \frac{\partial^{2} \psi}{\partial q_{3}^{2}}\right]\right\}  + V_{\lambda}\psi = E\psi,
\end{equation}
where
\begin{equation}\label{DLAP}
D(q)\psi = \displaystyle\sum_{i,j=1}^{2}\frac{1}{\sqrt{|\det(G)|}}\frac{\partial}{\partial q_{i}}\left(\sqrt{|\det(G)|}G^{ij}\frac{\partial \psi}{\partial q_{j}}\right).
\end{equation} 
Note that the wave function $\psi (q_1,q_2,q_3 )$ is defined in a three-dimensional neighborhood $\mathcal{N}$ of $p\in S$. From now on we assume the existence of a wave function $\chi$, that after the confinement process,  splits into a tangent ($\chi_{T}(q_{1},q_{2})$) and a normal ($\chi_{N}(q_{3})$) contribution, such that 
\begin{equation} 
\displaystyle \int |\chi|^{2}\rmd S\rmd q_{3} = \int \left( |\chi_{T}|^{2} \displaystyle \int |\chi_{N}|^{2}\rmd q_{3}\right)\sqrt{|g|}\rmd q_{1}\rmd q_{2},
\end{equation}
where the term within brackets is the probability density on the surface. In order to do this, let us consider  the area element  $\rmd S=\sqrt{|\det(g)|}\rmd q_{1}\rmd q_{2}$ with $g$ the induced metric in $S$ given by (\ref{indg}). After straightforward calculations, the volume element in $\mathcal{N}$,  given by
\begin{equation}
\rmd \mathcal{V}  = \left\vert\left\langle \frac{\partial R}{\partial q_{1}} \times_{1} \frac{\partial R}{\partial q_{2}}, \frac{\partial R}{\partial q_{3}} \right\rangle_{1}\right\vert \rmd q_{1}\,\rmd q_{2}\,\rmd q_{3},
\end{equation}
is found to be
$
\rmd \mathcal{V}= |f(q_1,q_2,q_3)| \rmd S \rmd q_3,
$
where
\begin{equation}\label{f}
f(q) = \varepsilon[1 + q_{3}(a_{11} + a_{22}) + q_{3}^{2}(a_{11}a_{22}-a_{12}a_{21})]\,.
\end{equation}
Then
$\int |\psi|^{2}\rmd \mathcal{V} = \displaystyle \int |\psi|^{2}\left|f\right|\rmd S\rmd q_{3} =  \displaystyle \int |\chi|^{2}\rmd S\rmd q_{3}$,  where we defined  $\chi = \psi \sqrt{|f|}$. So,
since $\sqrt{|\det(G)|} = |f|\sqrt{|\det(g)|}$, equation $\eqref{DLAP}$ becomes
\begin{equation}\label{D}
D\left( \frac{\chi}{\sqrt{|f|}}\right)  =  \displaystyle \sum_{i,j=1}^{2} \frac{1}{|f|\sqrt{|\det g|}}\frac{\partial}{\partial q_{i}}\Big[|f|\sqrt{|\det g|}\frac{G^{ij}}{\sqrt{|f|}}\times\Big( \frac{\partial \chi}{\partial q_{j}} -\frac{\chi}{2}\frac{\partial}{\partial q_{j}}\ln|f|\Big)\Big] .
\end{equation}

Since we are dealing with a thin layer, we assume that $q_3 \in (-\delta,\delta)$. Now taking the limit $\delta \rightarrow 0$, after substitution of (\ref{D}), we get
\begin{equation}
 \rmi\hbar \frac{\partial \chi}{\partial t} =-\frac{\hbar^{2}}{2m}\Delta_g\chi-\varepsilon \frac{\hbar^{2}}{2m} \left\{\left[\frac{\mbox{tr}(a_{ij})}{2}  \right]^{2} - \det(a_{ij}) \right\}\chi - \varepsilon \frac{\hbar^{2}}{2m}\frac{\partial^{2}\chi}{\partial q_{3}^{2}} + V_{\lambda}(q_{3})\chi.\label{eqschHK}
\end{equation}
We proceed now to separate the variables in (\ref{eqschHK}) writing $\chi$=$\chi_{N}(q_{1},q_{2},t)\,\chi_{T}(q_{3},t)$, which leads to 
\begin{equation}\label{eqschrN}
-\varepsilon\frac{\hbar^{2}}{2m}\frac{\partial^{2}\chi_{N}}{\partial q_{3}^{2}} +  V_{\lambda}(q_{3})\chi_{N} = \rmi\hbar \frac{\partial \chi_N}{\partial t}
\end{equation}
and
\begin{equation}\label{eqschrT}
 \frac{-\hbar^{2}}{2m}\Delta_g\chi_T-\varepsilon \frac{\hbar^{2}}{2m} \left\{\left[\frac{\mbox{tr}(a)}{2}  \right]^{2} - \det(a) \right\}\chi_{T}  = \rmi\hbar \frac{\partial \chi_T}{\partial t}. 
\end{equation}
Equations $\eqref{eqschrN}$ and $\eqref{eqschrT}$ are analogous to the ones obtained by da Costa \cite{da1981quantum}. Recall that, to assure the confinement of the particle to the surface, we inserted $\varepsilon$ in front of the potential in equation~\eqref{eq::DefViaLimVlambda}. In fact, this is a consequence of the causal character of the surface. For instance, if the surface is space-like, $q_3N$ is time-like ($\varepsilon=-1$) and therefore the first term in $\eqref{eqschrN}$ becomes positive, which is equivalent to having a negative mass in the usual Schr\"odinger equation in Euclidean space. Although particles with intrinsic negative masses are not known, negative effective masses appear in electronic metamaterials \cite{dragoman2007metamaterials}, Bose-Einstein condensates \cite{khamehchi2017negative} and optical excitations in semiconductors \cite{dhara2018anomalous}, for instance. Negative mass particles will be bound by repulsive potentials \cite{chen2018negative}, thus the necessity of inserting $\varepsilon$ in front of $V_{\lambda}$ as in equation (\ref{eq::DefViaLimVlambda}). On the other hand,  equation~\eqref{eqschrT} shows that the particle constrained to move on the surface $S$ is subjected to a geometry-induced potential which depends on the causal character of the surface.  

We remark that we are considering a mock 3D spacetime and  therefore real time (the parameter $t$ appearing in (\ref{eqsch})) is an external variable and is not mixed in with the other three coordinates. If we assume that the surface of interest is static (does not change its shape with time), the operators appearing on the left-hand side of (\ref{eqschrT}) do not depend on $t$, and we can assume the usual \emph{ansatz} $\chi_T(q_1,q_2,t)=\rme^{-\rmi Et/\hbar}\varphi (q_1,q_2)$ to obtain the time-independent Schr\"odinger equation as
\begin{equation}\label{eqscht}
-\frac{\hbar^{2}}{2m} \Delta_{g} \varphi + V_{S}(q_{1}, q_2)\varphi = E \varphi,
\end{equation}
and, using equations \eqref{H} and \eqref{K}, 
\begin{equation}
V_S= -\varepsilon \frac{\hslash^{2}}{2m} \left\{\left[\frac{\mbox{tr}(a)}{2}  \right]^{2} - \det(a) \right\} = - \frac{\hslash^{2}}{2m} \left(\varepsilon H^2-K\right). \label{VS}
\end{equation}
Here $V_S (q_1,q_2)$
is the geometry-induced potential associated with the confinement of the particle to the surface $S$. Besides, it is noteworthy that both the mean curvature and the causal character of the surface, which are extrinsic properties, and thus depend on how the surface is immersed, appear together. The following comment  by da Costa also applies: ``\emph{Strange as it may appear at first sight, this is not an unexpected result, since, independent  of how small the range of values assumed for $q_3$, the wave function always `moves' in a three-dimensional portion of the space, so that the particle is `aware' of the external properties of the limit surface $S$.}" (Da Costa \cite{da1981quantum}, p. 1984).
We mention that $H^2-\varepsilon K$ is related to the standard deviation of the normal curvatures seen as a random variable \cite{da2017moving} and, therefore, the particle sees the extrinsic geometry as long as the surface does not bend equally in all directions. In fact, $H^2-\varepsilon K$ is the square root of the discriminant of the characteristic polynomial of the shape operator  \cite{da2017moving}, which vanishes at an umbilic, and then it measures how much $S$ deviates from curving equally in all directions.

Equation (\ref{VS}) is our main result and, in order to gain more insight into its meaning,  we make applications to surfaces of revolution in the following section. (The problem of finding surfaces of revolution in $\mathbb{R}^3$ with a prescribed $H^2-K$ was solved in \cite{daSilva2017quantumVgip} in the context of a constrained dynamics, while the same problem for  surfaces of revolution in $\mathbb{R}_1^3$ is solved in \cite{da2017moving} for mathematical purposes only.)

 It is worth mentioning that, when compared to the constrained dynamics in the usual Euclidean space, the main difference between the geometry-induced potential given by (\ref{VS}) and the one obtained by da Costa \cite{da1981quantum} lies in the causal factor $\varepsilon$. On one hand, for any time-like surface ($\varepsilon=1$) in $\mathbb{R}^{3}_{1}$, the effective constrained dynamics will be subjected to  a $V_S$ which is formally identical to da Costa's original potential: 
\begin{equation}
V_S = -\frac{\hbar^2}{2m}(H^2-K)\Rightarrow\left\{\begin{array}{lcc}
V_S\leq0, &A_p& \mbox{ diagonalizable}\\
V_S>0, &A_p& \mbox{ non-diagonalizable}\\
\end{array}
\right..
\end{equation}
Observe, however, that the dynamics is not expected to be the same since the Laplacian on a time-like surface is no longer an elliptic operator. Indeed, it comes from a metric of Lorentzian signature $(-,+)$. In addition, note that unlike the usual geometry-induced potential, the $V_S$ above does not have always the same sign. More precisely, since $H^2-\varepsilon K$ is the discriminant of the characteristic polynomial of the shape operator $A$ \cite{lopez}, one has $V_S\leq0$ if $A$ is diagonalizable and $V_S\geq0$ otherwise. Finally, taking into account that a time-like direction in $S$ may be associated with an effective negative mass, even if $V_S$ has the same sign at all points of the surface, it acts attractively or repulsively along directions with distinct causal characters.

On the other hand, for any space-like surface ($\varepsilon=-1$) in $\mathbb{R}_1^3$, the Laplacian is an elliptic operator since it comes from a metric of Riemannian signature $(+,+)$. However, unlike the usual constrained dynamics in Euclidean space, where $V_{S}\leq0$, the geometry-induced potential associated with a space-like surface always acts repulsively. Indeed, since the shape operator of a space-like surface is always diagonalizable \cite{lopez}, we necessarily have $H^2-\varepsilon K\geq 0$ and, therefore,
\begin{equation}
\varepsilon=-1\Longrightarrow V_{S}=\frac{\hbar^2}{2m}(H^2+K)\geq 0.
\end{equation}
In short, for a space-like surface we have an effective dynamics which is Riemannian (i.e., the Laplacian is elliptic), but subjected to a repulsive geometry-induced potential, while for a time-like surface we have an effective dynamics which is ``semi-Riemannian" (i.e., the Laplacian is non-elliptic), but subjected to a geometry-induced potential which acts differently along directions with distinct causal characters.

\section{Applications to surfaces of revolution}

Since the induced metric from $\mathbb{R}^{3}_{1}$ on surfaces of revolution with either a space- or a time-like axis is diagonal, the separation of variables of (\ref{eqscht}) is straightforward. In other words, taking $\varphi(q_1,q_2)=\chi_1 (q_1)\chi_2 (q_2)$ it follows that
\begin{equation}\label{osc1}
-\frac{\rmd^2\chi_{1}}{\rmd q_{1}^{2}} = \frac{2mE_{1}}{\hbar^{2}}\chi_{1}
\end{equation}
and
\begin{equation}\label{eqq4}
-\frac{1}{\sqrt{|\det(g)|}}\frac{\rmd}{\rmd q_{2}}\left( \sqrt{|\det(g)|}\,g^{22}\,\frac{\rmd \chi_{2}}{\rmd q_{2}} \right)
-\left[ \left(\varepsilon H^{2} - K \right) 
+ \frac{2m}{\hbar^{2}}\left( E - g^{11}E_{1}\right)\right]\chi_{2}=0,
\end{equation}
where $E_1$ is the separation of variables constant. While \eqref{osc1} depends only on $q_1$, the sole dependence of \eqref{eqq4} is on $q_2$ due to the rotational invariance. Note that the domain of $q_1$ is not necessarily $[0,2\pi )$ as it would always be in $\mathbb{R}^3$. In case of a hyperbolic rotation the domain is $(-\infty , \infty )$, which leads to a continuum spectrum for equation~\eqref{osc1}. 

In the following subsections, we show that (\ref{eqq4}) reduces to an effective 1D dynamics along the profile curve subjected to a 1D effective potential $V_{\text{\it eff}}$. Besides the geometry-induced potential $V_S$, there is another contribution to $V_{\text{\it eff}}$ which can be attributed to the intrinsic geometry of the surface of revolution only. The effective potential $V_{\text{\it eff}}$ that will appear in equations (\ref{eqSeta}), (\ref{eqSSS4}), and (\ref{eqTSeta}) below, can be decomposed into two terms. A contribution of the form $\pm k_2^2/4$ that can be seen as a geometry-induced potential for a particle constrained to move along the profile curve (here $k_2$ is the curvature function of the profile curve) and another contribution acting as a centripetal potential due to the revolution. The same phenomenon can be observed for helicoidal and revolution surfaces in Euclidean space \cite{daSilva2017quantumVgip,atanasovPhysLettA2007}, but here we draw the reader's attention to the fact that, for a particle constrained to move along a curve of curvature $\kappa$ on a space-like plane, the effective constrained dynamics is
\begin{equation}\label{eq::1dConstDynOnAspacePlane}
-\frac{\hbar^2}{2m}\frac{\rmd^2\psi}{\rmd\, s^2}-\frac{\hbar^2}{2m}\frac{\kappa^2}{4}\psi= E\psi.
\end{equation}
While for a particle constrained to move along a curve on a time--like plane, the effective constrained dynamics is
\begin{equation}\label{eq::1dConstDynOnAtimePlane}
-\frac{\hbar^2}{2m}\frac{\rmd^2\psi}{\rmd\, s^2}-\varepsilon\frac{\hbar^2}{2m}\frac{\kappa^2}{4}\psi= E\psi,
\end{equation}
where $\varepsilon=+1\,(-1)$ for a time-like (space-like) curve (compare these two last equations with the Euclidean result \cite{da1981quantum}). In other words, unlike the Euclidean case, where the nature of the two contributions to $V_{\text{\it eff}}$ only depends on the quantum number associated with the angular momentum in the axis direction, in Minkowski space the nature of these contributions, i.e., whether they act attractively or repulsively, also depends on the causal character of the profile curve and of the corresponding surface of revolution.

\subsection{Schr\"odinger equation for a surface of revolution with  space-like axis and profile curve in a time-like plane}
Let $\alpha(q_{2})$ = $(u(q_{2}),0,v(q_{2}))$ be the profile curve  in the  plane $x_1x_3$, parametrized by its arc length. Then, $g_{11}=u^2$, $g_{22}=\eta$, $g_{12}=g_{21}=0$, and $\det{g} = \eta u^{2}$. This way, $g^{11}=\dfrac{1}{u^{2}}$, $g^{22}=\eta$, where $\eta=1$ if the generatrix is space-like or $\eta=-1$ if it is time-like. Substitution of this in $\eqref{eqq4}$ gives
\begin{equation}\label{eqST1}
\frac{\rmd^{2}\chi_{2}}{\rmd \,q_{2}^{2}} +\dfrac{u'}{u}\frac{\rmd \chi_{2}}{\rmd q_{2}} + \eta \left[(\varepsilon H^{2}-K) + \frac{2m}{\hbar^{2}}\left(E-\frac{E_{1}}{u^{2}}\right)  \right]\chi_{2}=0.
\end{equation}

Note that \eqref{eqST1} is of the form $\chi_{2}'' +V_1 (q_2)\chi_{2}' +V_2 (q_2)\chi_{2}=0$. 
By making   $\chi_2 (q_{2}) = y(q_{2})w(q_{2})$, where $w(q_{2})$ is a non-vanishing function, we get $$y'' + \left(2 \frac{w'}{w} + V_1 \right)y' + \left( \frac{w''}{w}+ V_1 \frac{w'}{w} + V_2 \right) y=0.$$ Choosing $2 w'/w + V_1=0$, one gets $w(q_2)=\rme^{-\frac{1}{2}P(q_2)}$, where $P(q_2)$ is a primitive of the function $V_1= u'/u$. It follows that $w=u^{-1/2}$ and, therefore,
$\chi_2 (q_{2}) = y(q_{2})u(q_{2})^{-1/2}
$. 
This puts \eqref{eqST1} in the form
\begin{equation}\label{eqST2}
 y'' + \eta\Big[(\varepsilon H^{2}-K) +\frac{2m}{\hbar^{2}}(E-\frac{E_{1}}{u^2})+\frac{\eta(u')^{2}}{4u^{2}}-\frac{\eta u''u}{2u^{2}}\Big]y=0.
\end{equation}

Noting that  $(u')^{2} = (v')^{2} -\eta$, $\dfrac{u''}{u}=k_{1}k_{2}$, and $\eta\varepsilon = -1$, since they have opposite signs, equation \eqref{eqST2} becomes
\begin{equation}\label{eqSeta}
     -\frac{1}{\eta}\frac{\rmd^{2}y}{\rmd q_{2}^{2}} + \left[\left(\frac{2mE_{1}}{\hbar^{2}}+ \dfrac{1}{4}\right)\dfrac{1}{u^{2}} +\eta\dfrac{k^{2}_{2}}{4}- \frac{2mE}{\hbar^{2}} \right]y=0.
\end{equation}
The $\eta$ in front of the second derivative emphasizes that, effectively, the particle moving along a time-like profile curve behaves as it had a negative mass. In addition, since we have here a space-like axis (hyperbolic rotation) the angular momentum in the axis direction is not quantized ($\ell\in\mathbb{R}$). The 1D effective potential reads
\begin{equation}
V_{\text{\it eff}} = \frac{\ell^2+1/4}{u^2}+\eta\,\frac{k_2^2}{4}\,,\ell\in\mathbb{R}.
\end{equation}
The term depending on $k_2$ corresponds to a confinement along the profile curve, see equation (\ref{eq::1dConstDynOnAtimePlane}). On the other hand, unlike the Euclidean space case, where the term depending on $\ell$ changes from centrifugal to anti-centrifugal for distinct  angular momentum quantum numbers \cite{atanasovPhysLettA2007}, this does not happen here.

\subsection{Schr\"odinger equation for a surface of revolution with  space-like axis and profile curve in a space-like plane}
Let us now consider the case of a space-like profile curve, parametrized by its arc length, in the plane $x_2x_3$, given by $\alpha(q_{2})$=$(0,u(q_{2}),v(q_{2}))$. Then, $g_{11}=-u^2$, $g_{22}=1$, $g_{12}=g_{21}=0$, and
$\det{g} = -u^{2}$. Therefore, $g^{11}=-1/u^{2}$, $g^{22}=1$. Substituting this into $\eqref{eqq4}$, we get
\begin{align}\label{eq1SSS}
 & \frac{\rmd^{2}\chi_{2}}{\rmd q_{2}^{2}} +\dfrac{u'}{u}\frac{\rmd \chi_{2}}{\rmd q_{2}} + \left[(\varepsilon H^{2}-K) + \frac{2m}{\hbar^{2}}\left(E+\frac{E_{1}}{u^{2}}\right)  \right]\chi_{2}=0.
\end{align}
Using the same trick as above results in
\begin{equation}\label{eq1SSS2}
  -y'' - \left[(\varepsilon H^{2}-K) +\frac{(u')^{2}-2u''u}{4u^{2}}\right]y=\frac{2m}{\hbar^{2}}(E+\frac{E_1}{u^2})y.
\end{equation}
Since the profile curve is in a space-like plane, $(u')^{2} = 1- (v')^{2}$ and $\left\langle N,N\right\rangle_{1}=\varepsilon = -1$, because $S$ is time-like. Furthermore, 
 $
 k_{1}k_{2} = \dfrac{-v'}{u}\left(v'u''-v''u' \right) = -\dfrac{u''}{u} .
 $
Thus, \eqref{eq1SSS2} becomes
\begin{equation}\label{eqSSS4}
  -\frac{\rmd^{2}y}{\rmd q_{2}^{2}} + \left[-\left(\frac{2mE_{1}}{\hbar^{2}}- \dfrac{1}{4}\right)\frac{1}{u^{2}}+ \dfrac{k^{2}_{2}}{4}-\frac{2mE}{\hbar^{2}} \right]y=0.
\end{equation}
Since we have a space-like axis, the angular momentum in the axis direction is not quantized ($\ell\in\mathbb{R}$) and the 1D effective potential reads
\begin{equation}
V_{\text{\it eff}} = -\frac{\ell^2-1/4}{u^2}+\frac{k_2^2}{4}\,,\ell\in\mathbb{R}.
\end{equation}
The term depending on $k_2$ corresponds to a confinement along the profile curve, Eq. (\ref{eq::1dConstDynOnAtimePlane}).

\subsection{Schr\"odinger equation for a surface of revolution with  time-like axis}
Next, we consider a profile curve  $\alpha(q_{2})=(u(q_{2}),0, v(q_{2}))$ in the plane $x_1x_3$ being rotated around the time-like axis $x_1$ and parametrized by its arc-length. It follows that $g_{11}=v^{2}$, $g_{22} = \eta$, $g_{12}=g_{21}=0$, and $\det{g}=\eta v^{2}$. Consequently, $g^{11}=1/v^{2}$, 
$g^{22} = \eta$ and, after substitution of these into $\eqref{eqq4}$, we have 
\begin{equation}\label{eqTT1}
\frac{\rmd^{2}\chi_{2}}{\rmd q_{2}^2} + 
 \frac{v'}{v}\frac{\rmd \chi_{2}}{\rmd q_{2}} + \eta\left[(\varepsilon H^2-K) + \frac{2m}{\hbar^{2}}(E 
 - \frac{E_{1}}{v^{2}})\right]\chi_{2} = 0.
\end{equation}
Now, using 
$\chi_2 (q_{2}) = y(q_{2})v(q_{2})^{-1/2} \label{subs2}
$, we get
\begin{equation}\label{eqTT2}
  y'' + \Big[\eta(\varepsilon H^2-K) +\eta \frac{2m}{\hbar^{2}}(E 
 -\frac{E_{1}}{v^{2}}) + \frac{(v')^{2}- 
 2v''v}{4v^{2}}\Big]y = 0. 
\end{equation}
Since $(v')^{2}$ = $(u')^{2} + \eta$,  $\eta\varepsilon = -1$, and 
$ 
k_{1}k_{2} = \frac{v''}{v},
$
it follows that
\begin{align}
     \eta(\varepsilon H^{2}-K) +\frac{(v')^{2}}{4v^{2}}-\frac{2v''v}{4v^{2}} &  = -\dfrac{k^{2}_{2}}{4} + \dfrac{\eta}{4v^{2}}\,.
\end{align}
Therefore $\eqref{eqTT2}$ is transformed into
\begin{equation}\label{eqTSeta}
     -\frac{1}{\eta}\frac{\rmd^{2}y}{\rmd q_{2}^{2}} + \left[\left(\frac{2mE_{1}}{\hbar^{2}}- \dfrac{1}{4}\right)\dfrac{1}{v^{2}} +\eta\dfrac{k^{2}_{2}}{4}- \frac{2mE}{\hbar^{2}} \right]y=0.
\end{equation}  
The $\eta$ in front of the second derivative is here to emphasize that, effectively, the particle moving along a time-like profile curve behaves as it were of negative mass. Observe that, unlike the case with a space-like rotation axis, here the angular momentum in the axis direction is quantized ($\ell\in\mathbb{Z}$). The 1D effective potential reads
\begin{equation}
V_{\text{\it eff}} = \frac{\ell^2-1/4}{u^2}+\eta\,\frac{k_2^2}{4}\,,\ell\in\mathbb{Z}.
\end{equation}
The term depending on $k_2$ corresponds to a confinement along the profile curve, Eq. (\ref{eq::1dConstDynOnAtimePlane}).

Equations  $\eqref{eqSeta}$, $\eqref{eqSSS4}$, and  $\eqref{eqTSeta}$, combined with  \eqref{osc1}, describe  the quantum motion of a particle constrained to surfaces of revolution in a three-dimensional space endowed with the Lorentz metric  $\mbox{diag}(-1,1,1)$. Note that, in all cases the equations  depend  on the  $g_{11}$ coefficient of the induced metric on the surface. 


\section{Examples: the one- and two-sheeted hyperboloids}

As examples, we consider the confinement of a quantum particle to one- and two-sheeted hyperboloids. Such surfaces of revolution in  $\mathbb{R}^{3}_{1}$ have a time-like axis and constant Gaussian curvature $+1$ in the one-sheeted case and   $-1$ in the two-sheeted case \cite{lopez}. They are, respectively, the pseudosphere $\mathbb{S}^{2}_{1}$ and the hyperbolic plane $\mathbb{H}^{2}$. In both cases we need  to solve \eqref{osc1} and \eqref{eqTSeta}. The first of these must be solved in the domain $q_1=[0,2\pi]$ with  periodic boundary conditions since the axis is time-like. It follows  that $\chi_{1}= \rme^{\rmi\ell q_{1}}$ and  $E_1=\ell^2\hbar^2/(2m)$, with $\ell$ integer. In addition, it is worth mentioning that both hyperboloids are totally umbilical surfaces and, consequently, $V_S$ does not contribute to the effective 1D dynamics along the profile curve, since $H^2-\varepsilon K \equiv 0 \Rightarrow V_S \equiv 0$. All the contribution to the effective dynamics along the profile curve comes from the intrinsic geometry. As will become clear below, unlike the usual Euclidean space, where the energy spectrum of a particle constrained to move in a sphere is discrete, in $\mathbb{R}_1^3$ both hyperboloids also present a continuous spectrum. This discrepancy between the spectra of totally umbilical surfaces in both $\mathbb{R}^3$ and $\mathbb{R}_1^3$ can be related to the fact that the sort of intrinsic geometries we may find in $\mathbb{R}_1^3$ differs from those found in Euclidean space: e.g., we may immerse the hyperbolic plane as a complete surface in $\mathbb{R}_1^3$, as a one sheet of the two-sheeted hyperboloid, but not in $\mathbb{R}^3$ (Hilbert Theorem). This shows that the difference between the sort of intrinsic geometries goes beyond the obvious fact that in $\mathbb{R}_1^3$ there are surfaces with non-Riemannian metrics but not in Euclidean space. In short, we hope the examples below can illustrate the special features associated with a quantum particle constrained to move on a surface of a Minkowskian ambient space.

Equation \eqref{eqTSeta} for both one- and two-sheeted hyperboloids becomes particular cases of the second  P\"oschl-Teller equation \cite{poschl1933bemerkungen}, 
\begin{equation}
\left\{ -\frac{\partial^2}{\partial r^2} + \alpha_1^2 \left[ \frac{\kappa (\kappa -1)}{\sinh^2 \alpha_1 r} - \frac{\lambda (\lambda +1)}{\cosh^2 \alpha_1 r}  \right]\right\} \psi =\frac{2ME}{\hbar^2}\psi  
.\label{PT}
\end{equation}

It is worth mentioning that in Euclidean space the 1D effective dynamics for a particle constrained to move on a sphere of radius $R$ can be written as \cite{daSilva2017quantumVgip}
\begin{equation}\label{eq::Eff1dDynOnEuclSphere}
-\frac{\rmd^2\psi}{\rmd\,s^2}+\left[-\frac{1}{4R^2}+\left(\ell^2-\frac{1}{4}\right)\frac{\csc^2(s/R)}{R^2}\right]\psi=\frac{2mE}{\hbar^2}\psi,
\end{equation}
where $s\in[0,\pi R]$ with boundary conditions $\vert\psi(0)\vert,\vert\psi(\pi)\vert<\infty$  and $\ell$ is the component of the angular momentum in the axis direction (up to a factor $-1/4$, $V_{\textit{eff}}$ above is a particular instance of the first P\"oschl-Teller equation \cite{poschl1933bemerkungen}, but with distinct boundary conditions).
Here, the profile curve reads $\alpha(s)=R(\sin(s/R),0,\cos(s/R))$ and it has curvature $\kappa=1/R$, which leads to a 1D geometry-induced potential $V_C$ satisfying $-2m\,V_C/\hbar^2=(2R)^{-2}$. The eigenstates of the Laplacian on the sphere are the well known spherical harmonics $Y_{n\ell}$ and the energy spectrum is
\begin{equation}
E_{n\ell} = \frac{\hbar^2}{2mR^2}\,n(n+1),\,n\in\mathbb{Z}\mbox{ and }\ell=-n,-(n-1),\dots,0,\dots,n-1,n.
\end{equation}

\subsection{One-sheeted hyperboloid}
Consider  the one-sheeted hyperboloid obtained by rotation of the curve parameterized by $\alpha(q_2) = \left(R\sinh(q_{2}/R),0,R\cosh(q_{2}/R)\right)$ around the time-like axis $x_1$. Such a surface is time-like since $\left\langle\alpha',\alpha'\right\rangle_{1} = -1$, and has principal curvatures $k_{2}(q_{2})=k_{1}(q_{2}) = -u'/v =  -1/R $. After substitution of $E_1=\ell^2\hbar^2/(2m)$, equation \eqref{eqTSeta} becomes then
\begin{equation}\label{eqTSeta2}
      \frac{\rmd^{2}y}{\rmd q_{2}^{2}} + \left[- \frac{1}{4R^2} +\left(\ell^2- \dfrac{1}{4}\right)\frac{\sech^2 (q_2/R)}{R^2}  \right]y=\frac{2mE}{\hbar^{2}}y ,
\end{equation}
which corresponds to $\kappa =0$ and $\lambda=  |\ell|-\frac{1}{2} $, since $\ell \in \mathbb{Z}$ and the solutions for  \eqref{PT} are valid only for $\lambda > \kappa$.  We assume boundary conditions $y=0$ when $q_2=\pm\infty$. Following Landau and Lifshitz \cite{landau2013quantum}, we find the spectrum 
\begin{equation}
E_{n\ell}=\frac{\hbar^2}{2mR^2}(n-|\ell|)(n-|\ell|+1) \geq 0 ,\,0\leq n < |\ell| - 1/2 \label{En1}
\end{equation}
 Since this condition cannot be fulfilled for $\ell=0$, this state is not included in  \eqref{En1} and therefore, it is not a bound state (a globally attractive potential in 1D  has at least one bound state \cite{buellAJP1995}). This can be explained by  the change in sign of the ``potential'' $\left(\ell^2- 1/4\right)\sech^2 (q_2)$, in \eqref{eqTSeta2}, which makes it repulsive. It is worth mentioning that (\ref{eqTSeta2}) also posses a continuous spectrum made of negative values \cite{landau2013quantum}.

The expression above suggests an infinite band of negative energy  states unbounded from below, reminiscent of the Dirac sea, and a discrete set of positive energy states, like those of a particle confined to a box in Euclidean space. This upside down spectrum is a consequence of the causal character  of the surface ($\eta = -1$) which changes the sign of the energy, see $\eqref{eqTSeta}$.

\subsection{Two-sheeted hyperboloid}
Consider now the two-sheeted hyperboloid obtained by rotation of the space-like curve  $\alpha(q_2) = \left(u(q_2),0,v(q_2)\right) = \left(R\cosh(q_{2}/R),0,R\sinh(q_{2}/R)\right)$ around the time-like axis $x_1$. This surface is space-like and has principal curvatures given by   $k_{2}(q_{2})=k_{1}(q_{2}) = -u'/v =  -1/R $. Since in this case $\eta = 1$, substitution of these data in  $\eqref{eqTSeta}$ leads to
\begin{align}\label{eqyeta2}
 &-\dfrac{\rmd^{2}y}{\rmd q_{2}^{2}} + \left[  \frac{1}{4 R^2} + \left(\ell^{2}-\frac{1}{4}\right)\frac{\csch^{2}(q_{2}/R)}{R^2}
 \right]y = \frac{2mE}{\hbar^2}y,
\end{align}
which, as  \eqref{eqTSeta2}, is also a particular case of the second P\"oschl-Teller equation \eqref{PT}, but with an effective potential globally repulsive for $\ell\not=0$. For $\ell=0$, there is an infinite potential well at $q_2=0$, while $V_{\textit{eff}}\sim 1/4R^2$ for $q_2\gg1$. Unlike the effective Schr\"odinger equation in the one-sheeted hyperboloid, the wave functions of the Laplacian operator acting on the two-sheeted hyperboloid, which is a model for the hyperbolic plane $\mathbb{H}^2(R)$, are all non-normalizable and the energy spectrum is continuous \cite{carinenaJMP2011}, as it happens for a free particle in an Euclidean plane. In particular, the wave functions are no longer in $L^2(\mathbb{H}^2(R))$. 
 
Note that the energy distribution obtained here is distinct when compared to the one of the previous example: the continuous part of the spectrum corresponds to positive values while there is no discrete energy level, exactly like the usual behavior in Euclidean space. This is not surprising since we have here a non-compact (infinite) space-like surface. Finally, notice here the formal similarity with the equation governing the effective dynamics (\ref{eq::Eff1dDynOnEuclSphere}) on an Euclidean sphere $\mathbb{S}^2(R)$. However, in $\mathbb{S}^2(R)$ the particle moves in a compact region and presents a spectrum that is both positive and discrete.

\section{Concluding remarks}
Motivated by the experimental realization of 3D Minkowski space $\mathbb{R}_1^3$ in hyperbolic metamaterials, we studied the Schr\"odinger equation for a particle constrained to a surface in such an environment. Due to the  anisotropy of $\mathbb{R}_1^3$, a wide range of surface types is possible, such as space-like, time-like, light-like or mixed type surfaces. For surfaces of revolution, the choice of the  axis, if time-like or space-like, for instance, determines whether one has an ordinary rotation or a hyperbolic one (the equivalent of a boost in spacetime). We followed the steps of da Costa \cite{da1981quantum} for the derivation of a quantum Hamiltonian describing the dynamics of a particle bound to a surface immersed in the three-dimensional space $\mathbb{R}_1^3$. Like da Costa, we found a geometry-induced potential arising from the immersion of the surface in $\mathbb{R}_1^3$. Our geometry-induced potential depends not only on the mean and Gaussian curvatures of the surface, as in the Euclidean case, but also on the causal character of the surface, as could be expected. As applications, we chose surfaces of revolution with space-like and time-like axes, and in each case a separable Schr\"odinger equation was obtained. We also provided three examples (particle in a box, one- and two-sheeted hyperboloids)  where the Schr\"odinger equation is exactly solvable and points to important differences in comparison with the dynamics in Euclidean space. It is worth mentioning the existence of
an alternative description of the constrained dynamics formalism in the context of a hyperbolic medium using
particles with negative effective masses in certain directions, but taking
into account an Euclidean background \cite{Filgueiras2018AlternativeMetaconfinement} instead of an effective Minkowski geometry, as done here. We also point out that our discussion of how to carry the interpretative postulates of Quantum Mechanics to $\mathbb{R}_1^3$ is absent in the approach of reference \cite{Filgueiras2018AlternativeMetaconfinement}. Finally, as perspectives, we mention the extension of the present work to more complex situations like a surface of revolution with a light-like axis, for instance, and surfaces with a curvature singularity as the compactified Milne universe model studied recently by one of us and coworkers \cite{figueiredo2017cosmology}. Besides, as commented in the Introduction, it is expected that the effect of the geometry-induced potential shall appear both in electronic and optical hyperbolic metamaterials. Therefore, we hope our results may be experimentally verifiable in the near future. 

\begin{acknowledgments}
We acknowledge partial financial support from CNPq, CAPES and FACEPE (Brazilian agencies).
\end{acknowledgments}

\end{document}